\documentclass[nofootinbib,amsmath,amssymb,10pt,eqsecnum, twocolumn]{revtex4-1}
\usepackage{graphicx, amsmath, amsmath, amssymb}
\usepackage{tabularx}
\usepackage[caption=false]{subfig}
\usepackage[breaklinks, colorlinks, citecolor=blue]{hyperref}
\usepackage{color}
\usepackage{float}
\usepackage{natbib,hyperref,ifthen}

\newcommand{\beq}{\begin{equation}}
\newcommand{\eeq}{\end{equation}}

\def\ee{\end{equation}}
\def\bea{\begin{eqnarray}}
\def\eea{\end{eqnarray}}
\def\bse{\begin{subequations}}

\newcolumntype{Y}{>{\centering\arraybackslash}X}

\begin{document}

\title{Improved Photometric Classification of Supernovae using Deep Learning}

\author{Adam Moss} \email{adam.moss@nottingham.ac.uk}
\affiliation{School of Physics \& Astronomy\\
University of Nottingham,
Nottingham, NG7 2RD, England}

\date{\today}

\begin{abstract}
   We present improved photometric supernovae classification using deep recurrent neural networks. The main improvements over previous work are (i) the introduction of a time gate in the recurrent cell that uses the observational time as an input; (ii) greatly increased data augmentation including time translation, addition of Gaussian noise and early truncation of the lightcurve. For post Supernovae Photometric Classification Challenge (SPCC) data, using a training fraction of $5.2\%$ (1103 supernovae) of a representational dataset, we obtain a type Ia vs. non type Ia classification accuracy of $93.2 \pm 0.1\%$, a Receiver Operating Characteristic curve AUC of $0.980 \pm 0.002$ and a SPCC figure-of-merit of $F_1=0.57 \pm 0.01$.  
 Using a representational dataset  of $50\%$ ($10,660$ supernovae),  we obtain a classification accuracy of  $96.6 \pm 0.1\%$, an AUC of $0.995 \pm 0.001$ and $F_1=0.76 \pm 0.01$. We found the non-representational training set of the SPCC resulted in a large degradation in performance due to a lack of faint supernovae, but this can be alleviated by the introduction of only a small number ($\sim 100$) of faint training samples. We also outline ways in which this could be achieved using unsupervised or semi-supervised domain adaptation. 
      \end{abstract}

\maketitle


\section{Introduction}

In the last few years there has been a revolution in the use of machine learning (ML), deep learning and
artificial intelligence to solve large classes of problems previously deemed intractable. Deep learning in particular has achieved state-of-the-art results in computer vision, speech recognition, natural language processing, search, and many more. It is a class of machine learning which aims to teach a computer an abstract representation of data. This representation is encoded by the weights of a neural network (NN), which consists of many layers of non-linear processing. There is an analogy between deep learning and the biological nervous system, which activates different neurons in the brain depending on the stimuli. While much of the conceptual and methodological basis was developed in the 80s and 90s, their extensive practical application became feasible only recently due to a combination of increased computational power and the availability of large data sets for NN training and analysis  (see~\cite{0483bd9444a348c8b59d54a190839ec9} for an overview).

Application of ML approaches is becoming widespread in the physical sciences. They are particularly suited
to the fields of astronomy and cosmology, which have increasingly large data sets (although the labelling of this
data for supervised learning can be problematic, and the training data is often not representative of the target). For example, the classification of transient objects is one of the biggest challenges facing the Large Synoptic Survey Telescope (LSST). These transient events include binary star systems, variable stars and
supernovae. The LSST will scan the sky at unprecedented depth, and is expected to produce up to
10 million transient alerts per night. 

 This was a primary motivation for the Supernova Photometric Classification Challenge (SPCC)~\cite{Kessler:2010wk,Kessler:2010qj}, developed to test supernovae  classification in photometric surveys. A spectroscopically confirmed training set was provided, along with an unseen test set. A variety of methods were used, with varying degrees of success, including template fitting and $\chi^2$ minimization. Later, ML methods were applied~\cite{Newling:2010bp, Karpenka:2012pm, 2015MNRAS.453.2848V, Lochner:2016hbn, 2018MNRAS.473.3969R}, all employing a two step process, where  features are first extracted from the lightcurve (e.g. by parametric fits, wavelet decomposition and principle component analysis) before classification. One of the advantages of deep learning is that it extracts  relevant features from data automatically.

Previously we have used  deep recurrent neural networks (RNNs) to classify supernovae~\cite{Charnock:2016ifh}. These are a
type of NN that can learn sequential data (e.g. time series and natural language) by forming connections along the
sequence, unlike feedforward NNs such as the multi-layer perceptron and convolutional neural networks. The hidden state of the RNN encodes an internal
representation of the input lightcurve, which can then be used for classification.

A variation of RNNs called Long Short Term Memory (LSTM)~\cite{LSTM} has a gating mechanism for the hidden state,
which enables longer sequences to be learnt by retaining `memory'. We found LSTM networks to perform better than standard RNNs, and were competitive with existing ML methods. A further advantage of RNNs is that they
can process sequences of arbitrary length so, given a transient alert, partial time series can be fed into the network
and classified to allow for potential follow up by other telescopes.

One issue with LSTMs is that they are not well suited for processing irregular time series. Recently, they
have been extended (P-LSTM~\cite{neil2016phased}) to include an additional time gate that updates the hidden state depending on the time (phase) of
 the observation, and have been shown to perform much better for the classification of asynchronous
sequences. This is advantageous as it is unlikely that lightcurves will be sampled uniformly due to gaps in observations. 

Another issue with our previous work was decreased performance on small training sets. Deep learning typically requires very large datasets, but this can be mitigated by augmenting the training data. In this work we greatly increase the effective size of the dataset by applying time translation symmetry, adding noise from the uncertainty in the flux measurement, and early truncation of the lightcurve.

The structure of the paper is as follows. In section~\ref{sec:network} we introduce the NN architecture used. In section~\ref{sec:data} we describe the data and augmentation methods. The classifier metrics are defined in section~\ref{sec:metrics} and results given in  section~\ref{sec:results}. Finally we give our conclusions in  section~\ref{sec:conclusions}.


\section{ Network Architecture } \label{sec:network}

\begin{figure}
\centering
\includegraphics[width=\columnwidth, angle=0]{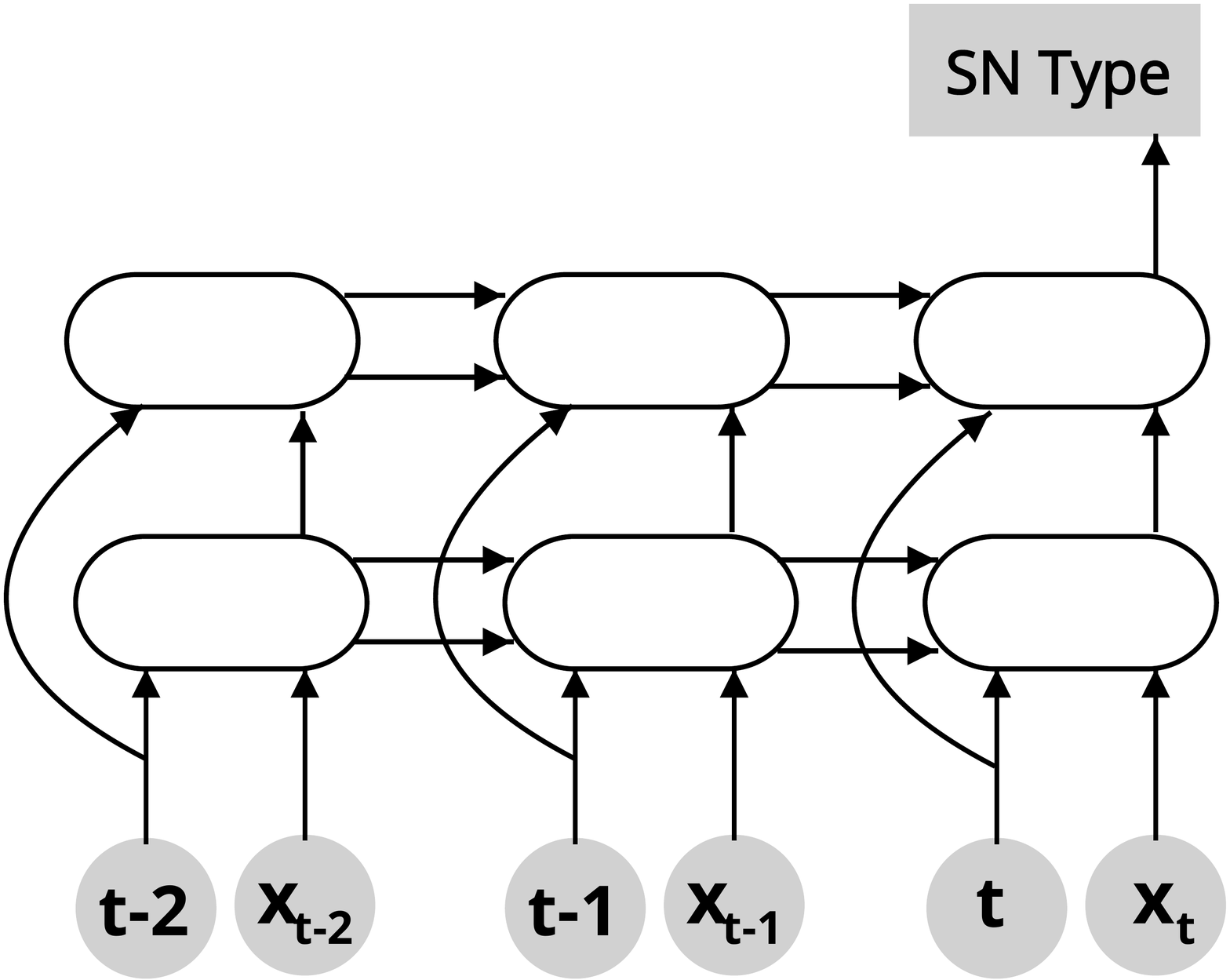}
\includegraphics[width=\columnwidth, angle=0]{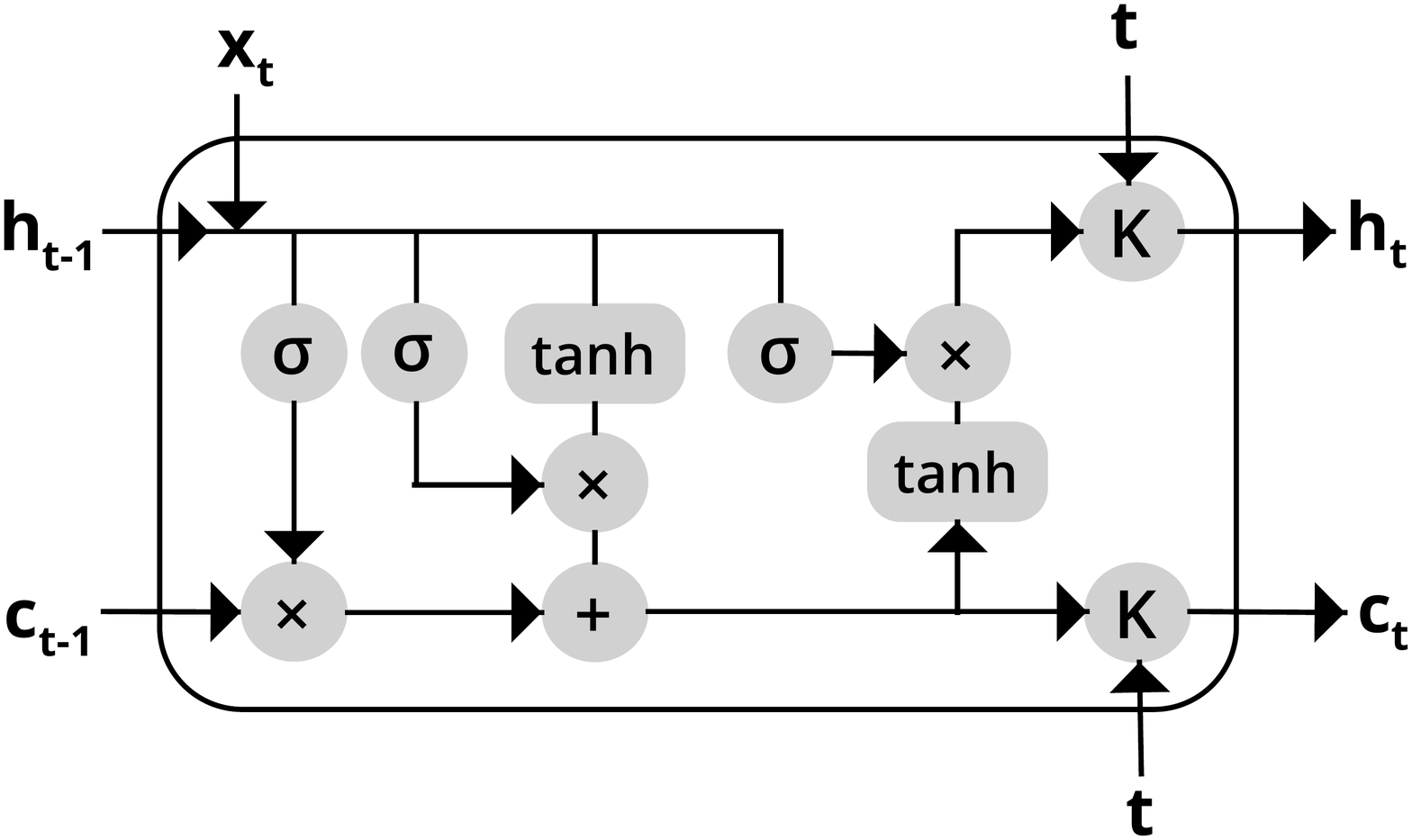}
\caption{\label{fig:phased} (Top) NN architecture used, with rounded boxes representing P-LSTM cells.  (Bottom) P-LSTM cell. The observational time $t$ and feature vector ${\bf x}_t$ are inputs to the network at each timestep.
 }
\end{figure}

The architecture of our network is shown in the upper panel of Fig.~\ref{fig:phased}, where the rounded boxes represent P-LSTM cells. The inputs consist of the observational time $t$ and a vector ${\bf x}_t$ at each sequential step.  This vector contains the flux in each band, along with any ancillary information available, such as host galaxy redshift and dust extinction. We use the same value at each sequential step for any time independent feature.  These are connected to a hidden
P-LSTM layer that encodes a representation of the input sequence.

Each P-LSTM cell contains a series of gates which control the propagation of the internal hidden state ${\bf h}_t$, whose dimension is equal to the number of hidden units. These are  the input gate ${\bf i}_t$, forget gate ${\bf f}_t$ and output gate ${\bf o}_t$. They take as inputs ${\bf x}_t$ and the previous hidden state ${\bf h}_{t-1}$,
\begin{eqnarray}
{\bf i}_t &=& \sigma \left( {\bf W}_{xi} {\bf x}_t + {\bf W}_{hi}  {\bf h}_{t-1}  + {\bf b}_i  \right)\,, \\ \nonumber
{\bf f}_t &=& \sigma \left( {\bf W}_{xf} {\bf x}_t + {\bf W}_{hf}  {\bf h}_{t-1}  + {\bf b}_f  \right)\,, \\ \nonumber
{\bf o}_t &=& \sigma \left( {\bf W}_{xo} {\bf x}_t + {\bf W}_{ho}  {\bf h}_{t-1}  + {\bf b}_o  \right)\,.
\end{eqnarray}
The sigmoid function $\sigma$ squashes the output in the range 0 to 1 and controls how much information to propagate from one step to the next.  The set of learnable weights ${\bf W}$ and biases ${\bf b}$ are the same for each cell in a single layer of the network,  initially taking random values and updated during training. 

The neuron also has a cell state ${\bf c}_t$, and a new candidate cell state ${\bf g}_t$ is obtained by passing ${\bf x}_t$ and ${\bf h}_{t-1}$ though a $\tanh$ activation function. These are then updated according to the gate outputs
\begin{eqnarray}
{\bf g}_t &=& \tanh \left( {\bf W}_{xg} {\bf x}_t + {\bf W}_{hg}  {\bf h}_{t-1}  + {\bf b}_g  \right)\,, \\ \nonumber
\tilde{\bf c}_t &=& {\bf f}_i \odot {\bf c}_{t-1} + {\bf i}_i \odot {\bf g}_t\,, \\ \nonumber
{\bf c}_t &=& {\bf k}_t \odot \tilde{\bf c}_t + (1-{\bf k}_t) \odot {\bf c}_{t-1} \,, \\ \nonumber
\tilde{\bf h}_t &=& {\bf o}_i \odot \tanh \tilde{\bf c}_t\,, \\ \nonumber
{\bf h}_t &=& {\bf k}_t \odot \tilde{\bf h}_t + (1-{\bf k}_t) \odot {\bf h}_{t-1} \,,
\end{eqnarray}
where $\odot$ denotes element wise multiplication. The time gate ${\bf k}_t$ also controls the flow of information and opens and closes periodically according to the input time~\cite{neil2016phased}. The period and phase of the time gate  are learnable parameter and are different for each hidden unit. The final structure of the P-LSTM cell is shown in the lower panel of Fig.~\ref{fig:phased}. 

The output ${\bf h}_{n}$ at the last step of the final hidden layer is then multiplied by a weight and the bias added,
\begin{equation}
{\bf z} = {\bf W}_{f} {\bf h}_n + {\bf b}_f\,.
\end{equation}
This is fed to a softmax layer, which takes the input ${\bf z}$ and returns
normalised, exponentiated outputs for each class label
$i$, $\exp(z_i)/\sum_i \exp(z_i)$.
The NN is trained by backpropagating
the error from the categorical cross-entropy loss between predictions and targets using the Adam optimiser~\cite{2014arXiv1412.6980K}.

The model is coded in TensorFlow~\footnote{\href{https://www.tensorflow.org/}{https://www.tensorflow.org/} } and is available on request from the author. All results in this paper can be reproduced by running scripts accompanying the code. 


\section{ Data} \label{sec:data}

In this paper we use post Supernovae Photometric Classification Challenge (SPCC)~\cite{Kessler:2010wk,Kessler:2010qj} data. The original dataset consisted of $\sim23,000$ simulated light curves, but during the challenge errors were discovered, resulting in a post challenge dataset of 21,319 supernovae\footnote{This is the SIMGEN\_PUBLIC\_DES dataset}.  These errors included an overestimation of the brightness of type Ia, and an underestimation of the brightness of non type Ia supernovae. It is important to note that results obtained from post SPCC data are not comparable to results from the challenge itself, as these errors can make it can make it easier to distinguish supernovae. For this reason we primarily compare our results to other methods using post SPCC data.

Each supernovae consists of a time series of flux measurements, with errors, in the $g, r, i, z$ bands, along with the position on the sky and dust extinction. Two separate challenges were given depending on whether the host galaxy redshift was available or not. 

\begin{figure*}
\centering
\includegraphics[width=\textwidth]{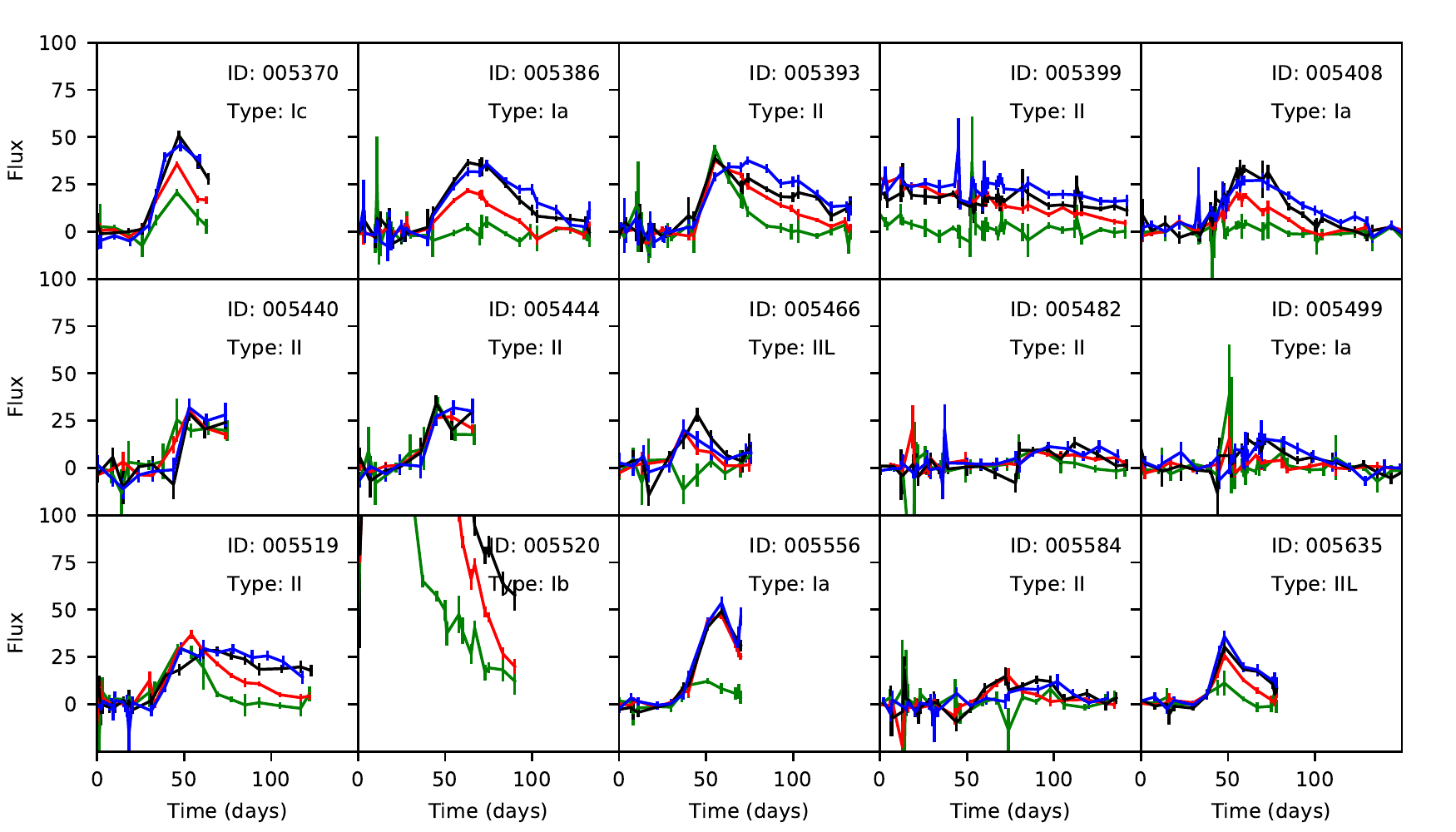}
\includegraphics[width=\textwidth]{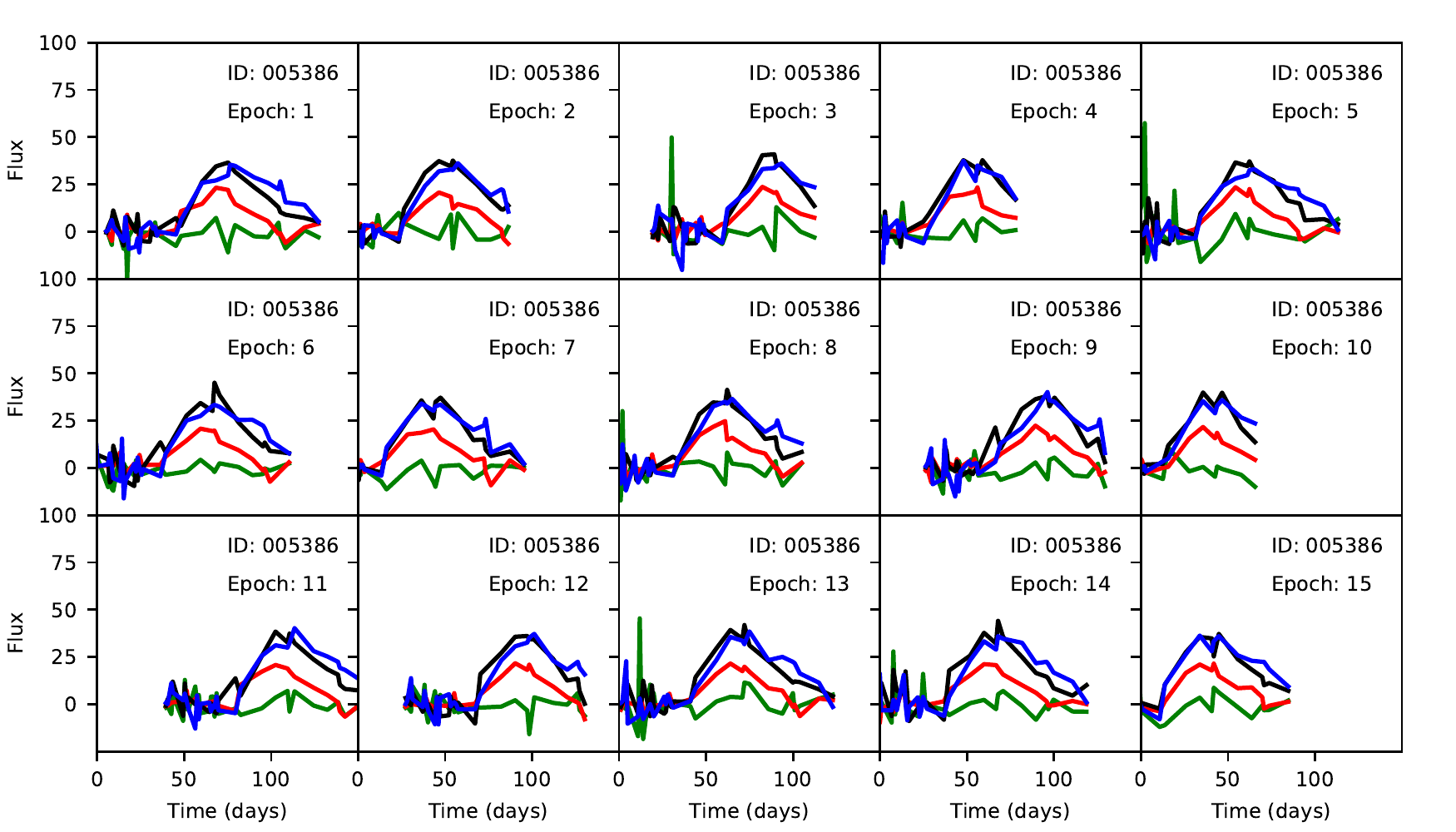}
\caption{\label{fig:lightcurves} (Top) Example light curves in the $g$ (green), $r$ (red), $i$ (black) and $z$ (blue) bands of the post SPCC dataset. (Bottom) Training examples per epoch of the type Ia supernovae SN005386. The lightcurves have random imputed missing values, time translation, added Gaussian noise and random early truncation. 
 }
\end{figure*}

The dataset consists of a total of 8 supernovae classes, types Ia, II (with sub-classes IIn, IIP, IIL) and Ibc (with sub-classes Ib and Ic). We consider these as three separate problems: the first being to categorise two classes (type Ia vs non type Ia), the second to categorise three classes (supernovae types Ia, II and Ibc) and the third to categorise all 8 sub-classes. An example set of light curves are shown in the top panel of Fig.~\ref{fig:lightcurves}. 

\begin{figure}[h]
\centering
\includegraphics[width=70mm]{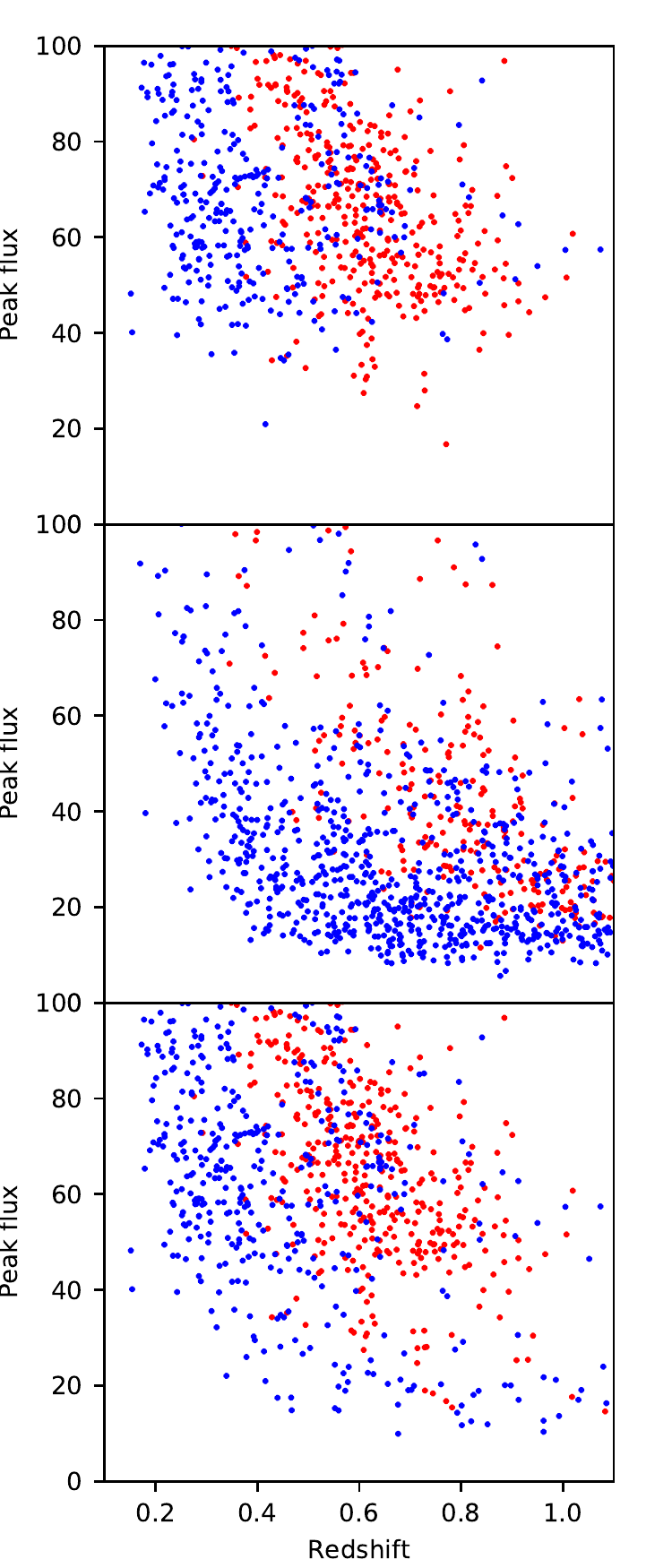}
\caption{\label{fig:trainingdata}  Peak $i$ band flux versus simulation redshift for (top) non-representative training data, (middle) representative training data of the same size, (bottom) non-representative training data + 100 representative samples. Type Ia supernovae are shown in red and non type Ia in blue. 
 }
\end{figure}

The training set consists of 1103 spectroscopically confirmed supernovae, corresponding to $5.2\%$ of the total samples. It is non-representative, containing both a class imbalance (the training set contains $50.7\%$ type Ia and the test set $23.9\%$ type Ia)  and very few faint samples. This is intended to mimic the limited spectroscopic resources of future surveys.  We show the training set in the top panel of Fig.~\ref{fig:trainingdata}, plotting the peak $i$ band flux versus the actual simulation redshift (this variable is not used as a input feature). 

Since we have all the class labels available, we can also test representative datasets of arbitrary size by randomly splitting into training and test subsets. In the middle panel of Fig.~\ref{fig:trainingdata} we show a representative sample of the same size. It is clear there are fewer type Ia supernovae and more faint objects. The full relative class breakdown for each case is shown in Tab.~\ref{tab:breakdown}. Deep learning performs better on representative training data, so we anticipate reduced performance in the non-representative case. 

\begin{table}[h]
\begin{tabularx}{\columnwidth}{Y Y Y} \hline \hline
Class  & Non-representative abdundance  (\%) & Representative  abundance (\%) \\ \hline \hline
Ia & 50.7 & 23.9 \\ 
II & 29.8 & 56.4 \\ 
IIn & 5.0 & 3.7 \\ 
IIP & 0.5 & 0.9 \\ 
IIL & 1.0 & 2.0 \\ 
Ibc & 1.4 & 1.2 \\ 
Ib & 6.4 & 6.7 \\ 
Ic & 5.3 & 5.2 \\ 
\end{tabularx}
\caption{\label{tab:breakdown} Supernovae class abundance for the non-representative and representative training sets.}
\end{table}

The input data is in the form of  time-ordered $g,r,i,z$ measurements, with one band observed per timestep. We therefore first perform data processing and group $g, r, i, z$ fluxes at a common time. This is possible as observations from each filter are clustered, with only occasional missing observations. The flexibility of the P-LSTM cell does allow for inputs of varying sampling rates if this isn't the case however. 

First, we normalize the time sequence to begin at day 0, rather than counting forwards and backwards from the maxima of the light curve. For observations less than $\sim1$ hour apart, we group the $g, r, i, z$ values into a single vector, ensuring there is at most one filter type in each group. If there is more than one filter type, we further subdivide the group using a finer time interval.  We take the time of the group to be the mean time of its individual observations, which is valid as the intervals are small compared to the characteristic time of the supernovae. 

Any missing values in the grouped observations are then imputed during training. Deep learning performs better with more data, so in this study we heavily augment data {\em online}, which eliminates the need to save many realizations to disk. Each supernovae  will then have a different data realisation in each epoch (a full pass over the training data).  

We apply the following online augmentations to training data 

\begin{enumerate}
\item Missing values are imputed to take a random uniform value between the previous and next valid measurement in that filter.
\item Gaussian noise is added to each observation according to the measurement uncertainty. 
\item The lightcurve has time translational symmetry, so the NN should learn to extract  features over time {\em intervals} rather than expecting them at a particular time. We therefore apply a random uniform time shift of the lightcurve between -40 and +40 days. 
\item Some of the observations are ended early, as shown in Fig.~\ref{fig:lightcurves}. In order to increase the number of training examples which end early, we randomly truncate the last $N$ days of observations, with $N$ taking a uniform value between 0 and 40, for any lightcurves exceeding 100 days.
\end{enumerate}
An example of these augmentations for each training epoch of SN005386 are shown in the lower panel of Fig.~\ref{fig:lightcurves}. 

For test data, the only augmentation we apply is imputing any missing values by taking the {\em average} value (rather than a random value) between the last and next valid observations. No additional noise, time translation or truncation is applied.


\section{Metrics} \label{sec:metrics}

The goal of the classifier is to determine the type of supernovae in the unseen test set. There are various metrics to assess the performance of classification tasks, the simplest being the accuracy.

For two class problems (e.g. type Ia versus non type Ia), the performance can be assessed in terms of the $2\times2$ confusion matrix. True positives (TP) are a correct positive prediction (correctly classifying as type Ia), false positives (FP) are a type 1 error (incorrectly classifying as type Ia), false negatives (FN) are a type 2 error (incorrectly classifying as a non type Ia) and true negatives (TN) are a correct negative prediction (correctly classifying as non type Ia). The actual number of type Ia supernovae is then TP + FN, and the actual number of non type Ia supernovae is TN + FP.

In terms of the confusion matrix, the accuracy, which is simply the ratio of the number of correct predictions to the total number of predictions, is
\begin{equation}
{\rm Accuracy} = \frac{{\rm TP + TN}}{{\rm TP + TN + FP + FN}}\,.
\end{equation}
The precision/purity is defined as 
\begin{equation}
{\rm Precision} = \frac{{\rm TP}}{{\rm TP}+{\rm FP}}\,,
\end{equation}
and the recall/completeness is defined as
\begin{equation}
\quad {\rm Recall} = \frac{{\rm TP}}{{\rm TP}+{\rm FN}}\,.
\end{equation}
A classifier with high precision but low recall would not make many predictions of type Ia (meaning it would miss lots), but of those it {\em does} make it is very precise. A classifier with low precision but high recall would make many predictions of type Ia, but most of these would be incorrect. 

The ideal classifier has both high precision and high recall. The $F_1$ score is usually defined as the harmonic mean of precision and recall, 
\begin{equation}
F_1 = \left(\frac{{\rm Precision}^{-1} + {\rm Recall}^{-1}}{2}\right)^{-1}\,.
\end{equation}
The SPCC defined  the figure-of-merit as
\begin{equation}
F_1 = \frac{1}{{\rm TP}+{\rm FN}} \frac{{\rm TP}^2}{{\rm TP}+3 \times {\rm FP}}\,,
\end{equation}
so false positives (incorrectly classifying a non type Ia supernovae as a type Ia) are penalised more heavily, as this is an expensive and time consuming mistake.

Finally, we also calculate  the Area Under the Curve (AUC) metric. The AUC is the area under the curve of the TP rate vs FP rate, as the threshold probability for classification is increased from 0 to 1. A perfect classifier has an AUC of 1, and a random classifier 0.5. 

For multi-class problems, the accuracy is again simply the ratio of the number of correct predictions to the total number of predictions. Rather than a $2 \times 2$ confusion matrix, we now have an $n \times n$ confusion matrix. The AUC can be calculated for each class vs the rest, and an unweighted or weighted average gives the final AUC score. 


\section{Results} \label{sec:results}

\begin{table*}[t!]
\begin{tabularx}{\textwidth}{Y | Y Y Y Y | Y Y Y Y Y } \hline \hline
Ref & Rep. size & Non-rep. size & SPCC data & Host z present& Accuracy (\%) & Precision (\%) & Recall (\%) & $F_1$ & AUC \\ \hline \hline
- &  1103  & - & Post & Yes & $ 93.2 \pm 0.1 $ &  $ 85.3 \pm 1.1 $ &  $ 86.7 \pm 1.2 $ &  $ 0.571 \pm 0.009 $ &  $ 0.980 \pm 0.002 $   \\
- & 1103  & - & Post & No &  $ 91.9 \pm 0.1 $ &  $ 82.6 \pm 1.3 $ &  $ 84.0 \pm 1.6 $ &  $ 0.514 \pm 0.009 $ &  $ 0.976 \pm 0.001 $   \\
- & 10,660  & -  & Post & Yes &  $ 96.6 \pm 0.1 $ &  $ 93.2 \pm 0.4 $ &  $ 92.2 \pm 0.8 $ &  $ 0.757 \pm 0.006 $ &  $ 0.995 \pm 0.001 $  \\
- & 10,660  & - & Post & No & $ 95.3 \pm 0.2 $ &  $ 89.5 \pm 0.7 $ &  $ 91.0 \pm 0.4 $ &  $ 0.674 \pm 0.013 $ &  $ 0.992 \pm 0.000 $  \\
- & -  & 1103 & Post & Yes & $ 57.5 \pm 4.3 $ &  $ 34.5 \pm 2.3 $ &  $ 97.3 \pm 0.6 $ &  $ 0.146 \pm 0.012 $ &  $ 0.601 \pm 0.065 $  \\
- & 50 & 1103 & Post & Yes &  $ 87.4 \pm 1.1 $ &  $ 68.5 \pm 4.1 $ &  $ 83.2 \pm 6.1 $ &  $ 0.349 \pm 0.025 $ &  $ 0.927 \pm 0.007 $ \\ 
- & 100 & 1103 & Post & Yes & $ 90.5 \pm 0.5 $ &  $ 76.3 \pm 2.2 $ &  $ 84.2 \pm 2.3 $ &  $ 0.436 \pm 0.017 $ &  $ 0.955 \pm 0.007 $  \\ 
- &  1256  & - & Original & Yes &  $ 94.3 \pm 0.3 $ &  $ 89.1 \pm 1.4 $ &  $ 91.8 \pm 1.2 $ &  $ 0.673 \pm 0.020 $ &  $ 0.983 \pm 0.002 $   \\
- &  -  & 1256 & Original & Yes &   $ 58.9 \pm 1.1 $ &  $ 38.9 \pm 0.7 $ &  $ 98.7 \pm 0.1 $ &  $ 0.173 \pm 0.004 $ &  $ 0.598 \pm 0.012 $  \\
- &  50  & 1256 & Original & Yes & $ 91.2 \pm 1.6 $ &  $ 78.6 \pm 4.5 $ &  $ 92.5 \pm 3.6 $ &  $ 0.512 \pm 0.047 $ &  $ 0.959 \pm 0.014 $  \\
- &  100  & 1256 & Original & Yes &  $ 92.3 \pm 0.6 $ &  $ 81.6 \pm 2.3 $ &  $ 91.5 \pm 2.6 $ &  $ 0.547 \pm 0.025 $ &  $ 0.966 \pm 0.003 $   \\
\hline \hline
~\cite{Newling:2010bp} & $\sim2000$  & - & Post & Yes &  &   &   & $\sim0.45$  &    \\
~\cite{Karpenka:2012pm} & 1045  & - & Post & Yes &  & 70  & 75  & 0.33  &    \\
~\cite{2015MNRAS.453.2848V} & $1256$  & - & Original & No &  & 84.3  & 85.7  & 0.55  &    \\
~\cite{2015MNRAS.453.2848V} & - & $1256$  & Original & No &  & 82.4  & 80.5  & 0.49  &    \\
~\cite{Lochner:2016hbn} & 1103  & - & Post & Yes &  & 90 & 85  &   & $ 0.984$   \\
~\cite{Lochner:2016hbn} & -  & 1103 & Post & Yes &  & - & -  &   & $\sim 0.9$   \\
~\cite{Charnock:2016ifh} & 1103  & - & Post & Yes & $   85.9 \pm  0.9  $ & $   72.4 \pm  0.4  $  & $  66.1 \pm 6.0  $  & $  0.31 \pm 0.03  $  & $ 0.910 \pm 0.012 $   \\
~\cite{Charnock:2016ifh} & 10,660  & - & Post & Yes & $   94.7 \pm  0.2  $ & $   87.3 \pm  0.8  $  & $  91.4 \pm 1.1  $  & $  0.64 \pm 0.01  $  & $ 0.986 \pm 0.001 $   \\
~\cite{2018MNRAS.473.3969R} & 1200 & $-$  & Original & Yes &  & 0.87  & 0.86   & 0.59  &  0.977  \\
~\cite{2018MNRAS.473.3969R} & - & $1217$  & Original & Yes &  &  &   &  &  0.961 \\
\end{tabularx}
\caption{\label{tab:metrics} Summary of results for  type Ia vs non type Ia classification. The network used is a 2 layer P-LSTM each with 128 hidden units. Rep. size and non-rep. size indicate the size of the representative and non-representative training sets respectively, and host z indicates if the host galaxy redshift was available or not. SPCC indicates whether the original or post SPCC dataset was  used. In the lower part of the table we compare against several ML studies.}
\end{table*}

Due to the increased data augmentation, we are able to use more complex networks that still generalize well to test data without over-fitting. This can be monitored during training by comparing the train and test losses -- a small training loss and high test loss indicates over-fitting. 

We tested several architectures, finding a 2 layer network with 128 hidden units each performed best, although we did not perform an exhaustive study of the  hyper-parameters. The network weights were trained in mini-batches of size 32 and each model was trained for 200 epochs.

We tested both representative and non-representative training samples of post SPCC data. In the representative case we performed 5  randomised train/test splits  to obtain
statistics of the classifier performance. In the non-representative case we only have a single training set, but again performed 5 runs to test convergence of the NN. Since the metrics vary from epoch-to-epoch, due to the stochastic nature of gradient descent, we average them over the last 10 epochs to obtain summary values. The results for type Ia vs non type Ia classification are shown in Tab.~\ref{tab:metrics}. 

For a representative training size of 1103 supernovae we obtain metrics substantially better than our previous analysis~\cite{Charnock:2016ifh}. The accuracy, for example, increases from 85.9\% to  93.2\% and the $F_1$ score from 0.31 to 0.57.   We have also compared against other ML approaches. A brief summary of their methodology is:
\begin{description}
\item[Newling~\cite{Newling:2010bp}] Parametric lightcurve fitting and kernel density estimation/boosting.
\item[Karpenka~\cite{Karpenka:2012pm}] Parametric lightcurve fitting and NN classification using a multi-layer perceptron.
\item[Varughese~\cite{2015MNRAS.453.2848V}] Wavelet decomposition of the lightcurve and ranked probability classification. The original SPCC dataset was  used in this study~\footnote{\href{http://www.hep.anl.gov/SNchallenge/}{This data is available at http://www.hep.anl.gov/SNchallenge/} }. 
\item[Lochner~\cite{Lochner:2016hbn}] Template fits, parametric fits and wavelet decomposition were investigated as feature extractors. K-nearest neighbours, multi-layer perceptrons, support vector machines and boosted decision trees were investigated as classifiers. 
\item[Charnock~\cite{Charnock:2016ifh}] LSTM recurrent neural networks.
\item[Revsbech~\cite{2018MNRAS.473.3969R}] Gaussian processes and random forests.
\end{description}

For a representative training sample of 1103 supernovae our method outperforms all other ML approaches apart from one: a combination of the model dependant SALT2~\cite{Guy:2007dv} (Spectral Adaptive Light curve Template 2) fit  and boosted decision tree classifier in ref.~\cite{Lochner:2016hbn}. Since  around 50\% of the simulated type Ia supernovae in the SPCC are generated using SALT2 templates, it is perhaps not surprising that this feature extractor performs well --  the results however are still comparable to ours.  
 
For a larger representative training set of 10,660 supernovae we again find improved metrics compared to our previous analysis, although the relative improvement is smaller. The accuracy, for example, increases from 94.7\% to 96.6\% and the $F_1$ score from 0.64 to 0.76. In all cases the inclusion of host galaxy redshift results in a small improvement in classifier performance.

In the non-representative case, our method suffers a large degradation in performance. The recall is high, meaning many (false) predictions of type Ia supernovae are made, with an accuracy of only 57.5\% and $F_1=0.15$. It is noticeable that the decrease in performance is worse than some other ML methods. In ref~\cite{Lochner:2016hbn}, for example, the best performing combination with non-representative data is a parametric lightcurve fit and boosted decision tree classifier, still achieved an AUC of $\sim 0.9$. 

\begin{figure}[h]
\centering
\includegraphics[width=70mm]{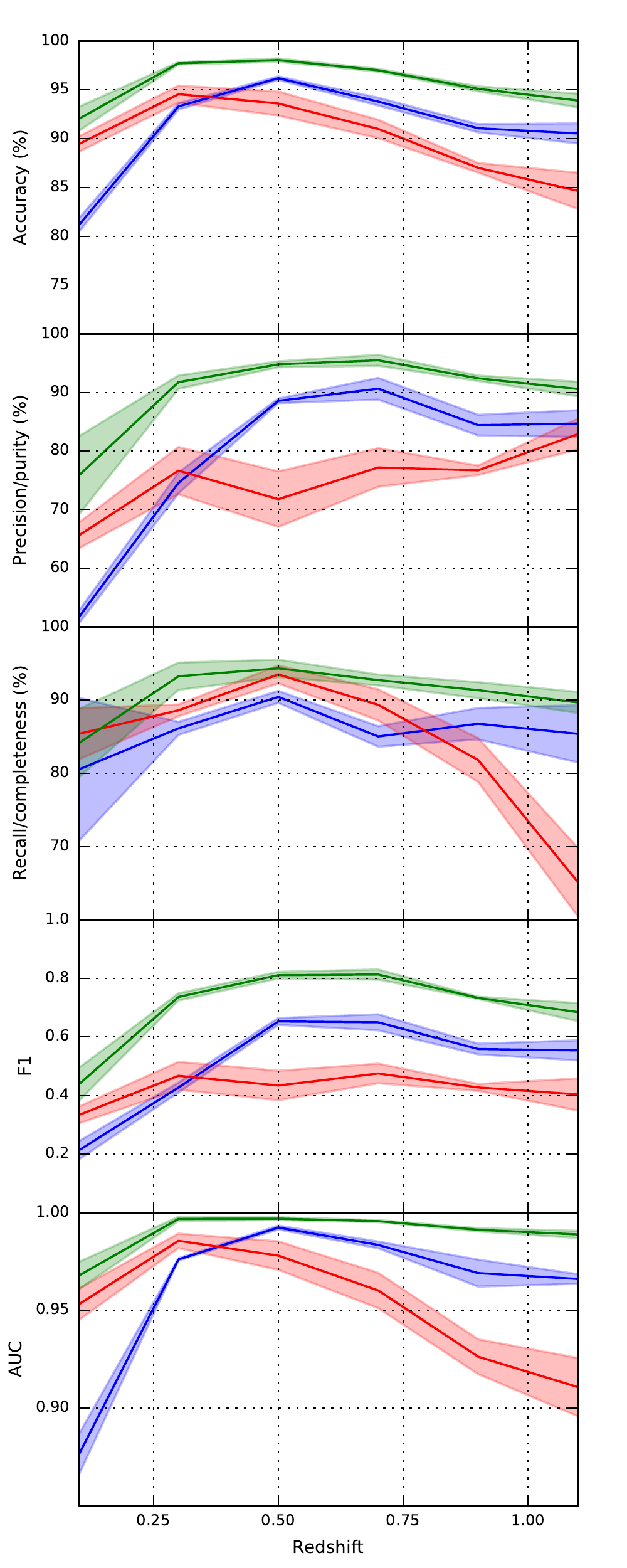}
\caption{\label{fig:metrics}  Classifier metrics as a function of simulation redshift for unseen test data. In blue we show results from a representative training sample of 1103 supernovae, in green a representative sample of 10,660 supernovae and in red the non-representative SPCC dataset with the addition of 100 representative samples. 
 }
\end{figure}

Non-representativeness is a common issue in deep learning when training on a {\em source} domain and applying to a new, but related, {\em target} domain with few or even zero samples. It can occur in many situations - for example labeled data might not be available in the target domain due to practical considerations, or the source domain may be simulations and the target domain real data. The goal of {\em domain adaptation} is to find a common representation space for the two domains. There have been recent advances in this field through the use of generative adversarial networks (GANs, see e.g.~\cite{2015arXiv150507818G}), which aims to align the domains through unsupervised or semi-supervised learning while retaining good performance on the source domain classification. 

As a precursor to the domain adaptation problem, we consider the simpler (but sub-optimal) method of supplementing the target domain with additional samples and training it jointly. This is shown in the lower panel of Fig.~\ref{fig:trainingdata}. In particular, we assume the non-representative dataset has an additional 50 or 100 representative samples.  We find that the classifier performance increases dramatically, the accuracy increasing from 57.5\% to 90.5\%, the $F_1$ score from 0.15 to 0.44 and the AUC from 0.601 to 0.955 in the latter case. It is clear that a limited number of faint samples helps to constrain the representation space of the target domain. Physically, we expect certain properties of supernovae lightcurves to transfer from bright to faint samples, so anticipate that by applying domain adaptation methods the joint representation can be improved even further. 

It is also instructive to show the classifier performance as a function of simulation redshift $z$. In Fig.~\ref{fig:metrics} we plot metrics as a function of $z$ by evaluating the test set in bins of $dz=0.2$ (to reiterate the redshift is not used as a feature during training).  Performance is worse at lower and higher redshift, where there are fewer training samples, and a maximum at $z \sim 0.5$.

We have also tested our method against the original SPCC dataset.  For a representative training sample of 1256 supernovae we obtained an increased accuracy of $94.3\%$ and $F_1=0.67$. This suggests the NN is efficient in exploiting errors in the original dataset. In the actual challenge itself a non-representative sample was used, and the highest entry had $F_1 \sim 0.4$, averaged across all redshift bins. We again see a large decrease  in performance using non-representative data, with an accuracy of only 58.9\% and $F_1=0.17$, but results improve dramatically with the addition of a  small number of faint supernovae. Supplementing the training with 100 representative samples leads to an improved accuracy of  92.3\% and $F_1=0.55$. Using only 50 samples gives $F_1=0.51$, still higher than any of the original challenge entries.

\begin{figure}[h]
\vspace*{0.5cm}
\centering
\includegraphics[width=88mm]{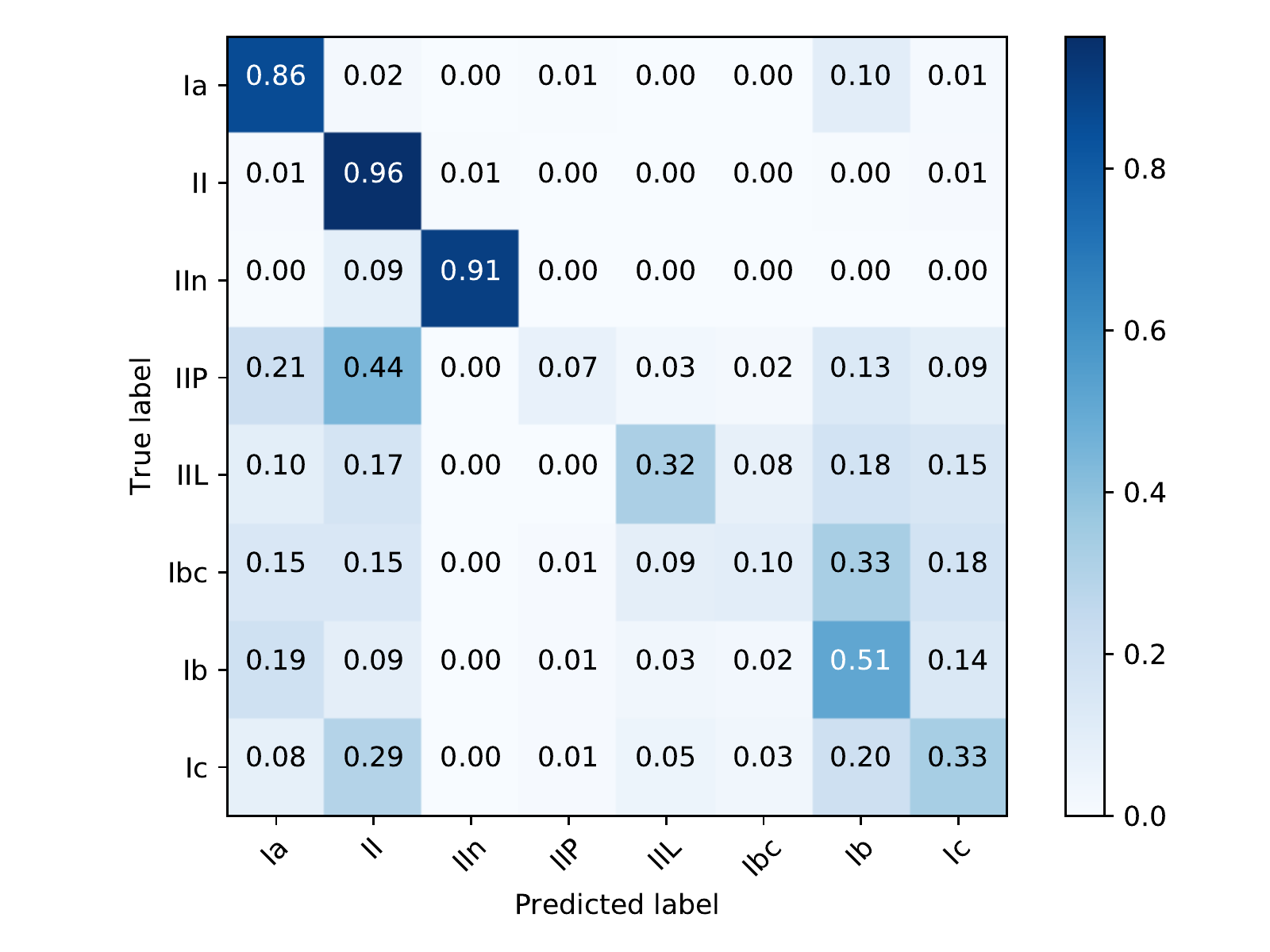}
\caption{\label{fig:confusion}  Normalized confusion matrix for the full eight class classification problem.
 }
\end{figure}

For the more difficult task of identifying the precise class of supernovae, we again find substantially improved results over our earlier work. For the three class problem (supernovae types Ia, II and Ibc) and a representative training sample of 1103 supernovae (post SPCC data), the accuracy improves from 78.1\% to $88.3 \pm 0.1 \%$  and the AUC from 0.868 to $0.970 \pm 0.001$. Previously, we did not attempt the full 8 class problem, but now obtain an accuracy of $83.4 \pm 0.3 \%$ and an AUC $0.968 \pm 0.002$. The confusion matrix is shown in Fig.~\ref{fig:confusion} for the full challenge, normalized such that the sum of the predicted labels is unity. The network is efficient at distinguishing between types Ia, II, and IIn, but is poorer at identifying Ib, even though it is the third most prevalent type in the training set. It fails to identify IIP completely, although this is not surprising given there are only $\sim 10$ training samples.


\section{Conclusions} \label{sec:conclusions}

We have presented significantly improved photometric classification of supernovae using deep learning. All previous ML approaches have used a two step process, first extracting features before classification, whereas our approach is model independent and lightcurves are used directly as inputs. For representative SPCC data we achieve state-of-the-art classification results. 

Neural networks are extremely good function approximators, forming an efficient internal representation of the inputs. However, this  representation space can become unconstrained when given related but atypical examples, compared to the data trained on. Our method performs poorly on the non-representative SPCC sample, but this can be alleviated with very few ($\sim 100$) representative samples with performance approaching the fully representative case. This suggests that future surveys should dedicate at least some of their time to obtain a small sample of spectroscopically confirmed faint objects.

There are several possibilities for future work. Recently, there have been many advances in the field of domain adaptation in deep learning, using both fully unsupervised and semi-supervised methods. Ref.~\cite{2017arXiv171102536M}, for example, suggests strategies for {\em few-shot learning}, where domains can be aligned with only a few (as low as one) examples per class in the non-representative target domain. 

The classification of transient events is a huge challenge facing the LSST. To prepare for this, the Photometric LSST Astronomical Time-series Classification Challenge (PLAsTiCC) has recently been released~\cite{Malz:2018zlf}, consisting of realistic simulated data containing a wide variety of transients (around 20 types).  To make the problem more realistic, the training sample is biased and not all types of transient events are included. This makes the problem even more challenging than the SPCC, but given the promising results we report here it will be interesting to apply our methods and develop new domain adaptation techniques for this problem. 

\section*{Acknowledgements}

We appreciate helpful conversations with Steven Bamford, Simon Dye and Juan Garrahan. A. M. is supported by a Royal Society University Research Fellowship.

\bibliography{bib}

\end{document}